\documentclass[prl,aps,twocolumn,superscriptaddress,showpacs]{revtex4}

\usepackage{graphicx}

\begin{document}

\title{Spin resonance in the d-wave superconductor CeCoIn$_{5}$}

\author{C. Stock}
\affiliation{Department of Physics and Astronomy, The Johns Hopkins University, Baltimore, Maryland USA 21218}
\author{C. Broholm}
\affiliation{Department of Physics and Astronomy, The Johns Hopkins University, Baltimore, Maryland USA 21218} 
\affiliation{NIST Center for Neutron Research, Gaithersburg, Maryland USA 20899}
\author{J. Hudis}
\affiliation{Department of Physics and Astronomy, The Johns Hopkins University, Baltimore, Maryland USA 21218}
\author{H. J. Kang}
\affiliation{NIST Center for Neutron Research, Gaithersburg, Maryland USA 20899}
\author{C. Petrovic}
\affiliation{Condensed Matter Physics, Brookhaven National Laboratory, Upton, New York, USA 11973}

\date{\today}

\begin{abstract}

Neutron scattering is used to probe antiferromagnetic spin fluctuations in the d-wave heavy fermion superconductor CeCoIn$_{5}$ (T$_{c}$=2.3 K). Superconductivity develops from a state with slow ($\hbar\Gamma$=0.3 $\pm$ 0.15 meV) commensurate (${\bf{Q_0}}$=(1/2,1/2,1/2)) antiferromagnetic spin fluctuations and nearly isotropic spin correlations.  The characteristic wavevector in CeCoIn$_{5}$ is the same as CeIn$_{3}$ but differs from the incommensurate wavevector measured in antiferromagnetically ordered CeRhIn$_{5}$.  A sharp spin resonance ($\hbar\Gamma<0.07$~meV) at $\hbar \omega$ = 0.60 $\pm$ 0.03 meV develops in the superconducting state removing spectral weight from low-energy transfers.   The presence of a resonance peak is indicative of strong coupling between f-electron magnetism and superconductivity and consistent with a d-wave gap order parameter satisfying $\Delta({\bf q+Q_0})=-\Delta({\bf q})$.

\end{abstract}


\maketitle

	While magnetism and s-wave superconductivity are thought to be incompatible, spin fluctuations may actually mediate superconductivity with other order parameter symmetries such as a d-wave gap symmetry.~\cite{Moriya00:49} A possible indication of this comes from the suggested observation of a spin resonance in the superconducting state of various high temperature superconductors. Intricacies of the cuprates, including the electronic inhomogeneity and the complex phase diagrams, complicates the interpretation so it is of
interest to examine related phenomena in chemically distinct materials which are homogeneous and lack electronic and structural disorder. In this letter we report the observation of a spin resonance in the CeCoIn$_{5}$ superconductor, a rare earth, and chemically homogeneous d-wave superconductor.

	$\rm CeCoIn_5$ is a superconductor with the highest recorded transition temperature thus far ($T_c=2.3$ K) for a heavy fermion material.\cite{Petrovic01:13} The compound is part of the Ce$T\rm In_5$ (T=Rh, Ir, and Co) family of inter-metallics featuring both antiferromagnetism (AFM) and superconductivity (SC). The tetragonal $\rm HoCoGa_5$ type crystal structure of these systems is built from alternating layers of $\rm CeIn_3$ and $T\rm In_2$ stacked along the [001] direction. The presence of two-dimensional planes of magnetic Ce$^{3+}$ ions links these materials to the SC cuprates and other materials where AFM and SC coexist.~\cite{Birgeneau06:75} The quasi two-dimensional nature is reflected in de Haas-van Alphen measurements which show that the Fermi surface includes nearly cylindrical surfaces and the largely two-dimensional nature of the electronic properties has been suggested as a contributing factor to the high superconducting transition temperature.~\cite{Monthoux02:66} This contrasts with AFM ordered $\rm CeRhIn_5$ where the Fermi surface exhibits a stronger three dimensional character.~\cite{Hall01:64,Hall01_2:64} The detected cyclotron masses of 5-87 $m_0$ are extremely large with the 4f electrons contributing greatly to the Fermi surface.~\cite{Settai01:13} While node locations within the basal plane remain under investigation~\cite{Izawa01:87,Aoki04:16}, it is clear that the SC state has d-wave symmetry.

	The experiments were conducted on the SPINS cold neutron spectrometer at the NIST Center for Neutron Research. A variable vertically focusing graphite (002) monochromator and a horizontally focused graphite (002) analyzer subtending a solid angle of $0.021$~sr with 11$^{\circ}$ horizontal acceptance were used. Energy transfer, $\hbar\omega= E_i-E_f$, was defined by fixing the final energy to 3.7 meV and varying the incident energy. Cooled Be and BeO filters were placed before and after the sample respectively. The sample consisted of $\sim$300 $\rm CeCoIn_5$ crystals with a total mass of 5 grams co-aligned on a series of aluminium plates so reflections of the form $(HHL)$ lay within the horizontal scattering plane. The room temperature lattice constants were measured to be $a=b=4.60$~\AA, and $c=7.51$~\AA. The crystals were secured to the mount using Fomblin-Y mechanical pump oil and covered by a thin aluminium plate. The combined mosaic spread was measured to be $\sim$ 3$^{\circ}$ and $\sim$ 5$^{\circ}$ for rotations around the $(1\overline{1}0)$ and $(001)$ directions respectively. The magnetic intensity was normalized using acoustic phonon scattering measured at $T=80$ K near ${\bf{Q}}=(111)$ through a constant $\hbar\omega=0.5$~meV scan.

\begin{figure}[t]
\includegraphics[width=75mm]{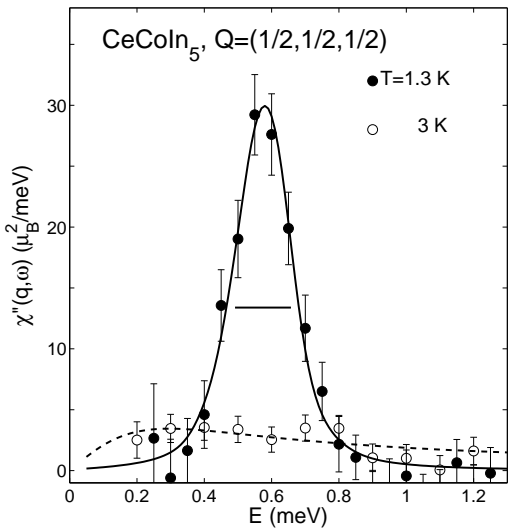}
\caption{The imaginary part of the dynamic susceptibility at ${\bf{Q}}=(\frac{1}{2}\frac{1}{2}\frac{1}{2})$ is plotted in the
normal (3 K) and in the superconducting (1.35 K) states. A background taken at {\bf{Q}}=(0.3,0.3,0.5) and {\bf{Q}}=(0.7,0.7,0.5) was subtracted. The horizontal bar is the resolution width.} \label{compare_temps}
\end{figure}

	The magnetic excitation spectrum at ${\bf{Q}}=(\frac{1}{2}\frac{1}{2}\frac{1}{2})$ in the normal state (T=3 K) and in the SC state (T=1.35 K) is plotted in Fig. \ref{compare_temps}. The normal state spectrum is featureless over the 1 meV energy range probed. The dashed line through these data show a Lorentzian response function, $\chi''({\bf{Q}},\omega) = \chi^\prime_{Q}\Gamma\omega/(\Gamma^2 + \omega^2)$, with a relaxation rate $\hbar\Gamma$=0.3 $\pm$ 0.15 meV. This contrasts with the excitation spectrum in the SC state which displays a sharp peak for $\hbar\omega_{0}=0.60\pm 0.03$~meV with a relaxation rate $\hbar\Gamma<0.07$~meV.

\begin{figure}[t]
\includegraphics[width=70mm]{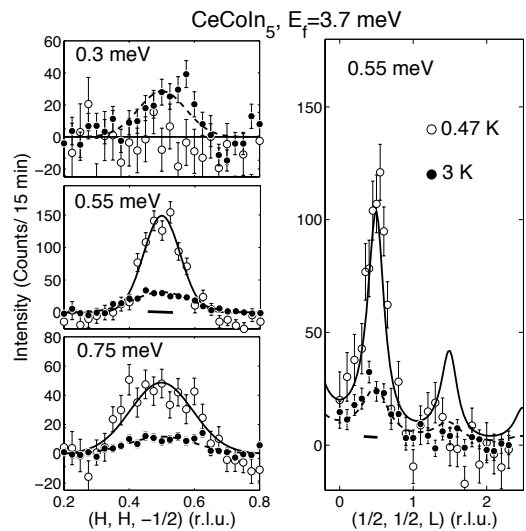}
\caption{Constant energy scans along the $(HH\frac{\overline{1}}{2})$ and $(\frac{1}{2}\frac{1}{2}L)$ directions in the normal and superconducting states. A featureless background measured at 15 K has been subtracted from the data. The horizontal bars represent the resolution width. The data has been corrected for absorption effects.} \label{constant_E}
\end{figure}

    The momentum dependence of the magnetic neutron scattering cross section is plotted in Fig. \ref{constant_E}. The left hand panel shows scans along the $(HH\frac{\overline{1}}{2})$ direction in the normal and SC states at several different values of energy transfer. The fits are to a single Gaussian function giving a dynamic correlation length (defined as the inverse of the half-width at half maximum) $\xi_{ab} = 9.6 \pm 1.0$ \AA\ at $\hbar\omega$=0.55 meV. In both the normal and SC phases, the magnetic scattering is peaked at ${\bf{Q}}=(\frac{1}{2}\frac{1}{2}\frac{1}{2})$ as for cubic CeIn$_{3}$~\cite{Knafo03:15} indicating nearest neighbor AFM
correlations within and between $a-b$ planes. 

    The right hand panel of Fig. \ref{constant_E} shows the wavevector dependence of the magnetic scattering along the $(\frac{1}{2}\frac{1}{2}L)$ direction. The solid line indicates a fit to $I({\bf Q})\propto f(Q)^{2}(1-({\bf {\hat{Q}\cdot \hat{c}}})^2) \sinh(c/\xi_c)/[\cosh(c/\xi_c)+\cos({\bf Q}\cdot{\bf
c})]$ which represents short-range AFM correlated Ce$^{3+}$ moments polarized along the [001] direction with a dynamic correlation length $\xi_c =6.5 \pm 0.9$ \AA\ at $\hbar \omega$ =0.55 meV. The squared form factor, $f(Q)^{2}$, was taken from previous measurements which agree with calculations by Blume \textit{et al.}~\cite{Blume62:37}   The second factor, $(1-({\bf {\hat{Q}\cdot \hat{c}}})^2)$, indicates spin fluctuations polarized along  ${\bf c}$ which is consistent with the anisotropic susceptibility. In contrast to the Fermi surface which is highly two-dimensional, the ratio of dynamic spin correlation lengths is only $\xi_{ab}/\xi_c$=1.5$\pm$ 0.4. 

    Commensurate AFM correlations distinguish CeCoIn$_{5}$ from CeRhIn$_{5}$, which develops AFM long range order with ${\bf{Q}}=(\frac{1}{2}\frac{1}{2}0.297)$.~\cite{Bao00:62} This is consistent with current understanding of the Fermi surfaces for these materials.~\cite{Hall01:64} While there is considerable modulation and potential for nesting along ${\bf c}$ for CeRhIn$_{5}$, the Fermi surface for CeCoIn$_{5}$ is comprised of cylinders with a large effective mass and little modulation along ${\bf c}$. It is interesting to note that both commensurate and incommensurate order has been observed in CeRh$_{1-x}$Co$_{x}$In$_{5}$ ($x$=0.4) and for similar concentrations in CeRh$_{1-x}$Ir$_{x}$In$_{5}$. AFM order and SC coexist in both compounds\cite{Yokoyama06:75,Christianson05:95} indicating a close connection between commensurate AFM spin fluctuations and SC.

\begin{figure}[t]
\includegraphics[width=80mm]{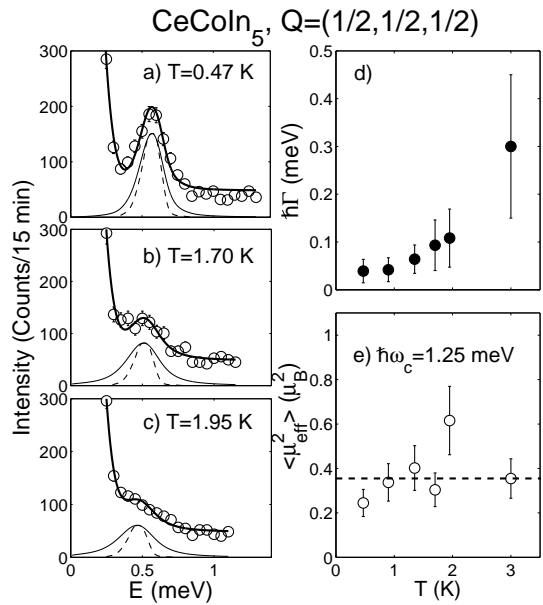}
\caption{(a)-(c) Constant-${\bf{Q}}=(\frac{1}{2}\frac{1}{2}\frac{1}{2})$ scans for $T<T_c=2.3$~K in $\rm CeCoIn_5$. The dashed line is the resolution function measured at $\hbar\omega$ = 0 and scaled to have the calculated width for finite energy transfers.(d) Half Width at Half Maximum for inelastic peak from fits described in text. (e) $\bf Q$ and $\hbar\omega$ integrated intensity below an energy cutoff $\hbar\omega_c=1.3$~meV.\label{constant_Q}}
\end{figure}

    Fig. \ref{constant_Q} shows three constant-Q scans at ${\bf{Q}}=(\frac{1}{2}\frac{1}{2}\frac{1}{2})$ for $T<T_c$. The solid lines are fits to a damped harmonic oscillator response function related to scattering through the fluctuation dissipation theorem. The fitting parameters include the staggered susceptibility $\chi'({\bf{q}})$, the resonance frequency $\omega_{0}$, and the relaxation rate, $\Gamma$. For the fits in Fig. \ref{constant_Q}, we fixed the background to be that measured at 15 K where the magnetic scattering is no longer peaked in momentum transfer. The fit provides a good description of the data at all temperatures with a low $T$ peak that broadens and softens on heating. Fig.~\ref{constant_Q} (d) shows  $\hbar\Gamma$ versus temperature. As $T$ approaches $T_{c}$, the peak broadens substantially in energy, less in momentum, indicating a decrease in life-time of collective spin fluctuations as the SC gap closes.

	The total moment sum-rule provides a measure of the fluctuating moment for $E<\hbar\omega_c$:

\begin{eqnarray}
\label{integrate} \nonumber <\mu_{\rm
eff}^{2}>_{\omega_{c}}=\frac{\hbar}{\mu_B^2\pi}\int_{0}^{\omega_{c}} d\omega
\int_{\Omega_{\bf q}} \frac{{\rm d}^{3}{\bf q}}{\Omega_{\bf q}}
\coth(\frac{1}{2}\beta\hbar\omega) {\rm Tr}\{\chi'' ({\bf
q},\omega)\} \nonumber
\end{eqnarray} 

\noindent Fig.~\ref{constant_Q} (e) shows that this quantity is approximately $T-$ independent in the temperature range probed. This indicates that the spectral weight under the resonance peak predominantly comes from low-energies as confirmed through the constant energy scan at $\hbar \omega$=0.3 meV illustrated in Fig. \ref{constant_E}. We note that $\langle \mu_{\rm eff}^2\rangle_{\omega_c}$ is comparable to that obtained in the elastic channel of AFM ordered CeIn$_{3}$ and CeRhIn$_{5}$.~\cite{Lawrence80:22,Bao00:62}

\begin{figure}[t]
\includegraphics[width=55mm]{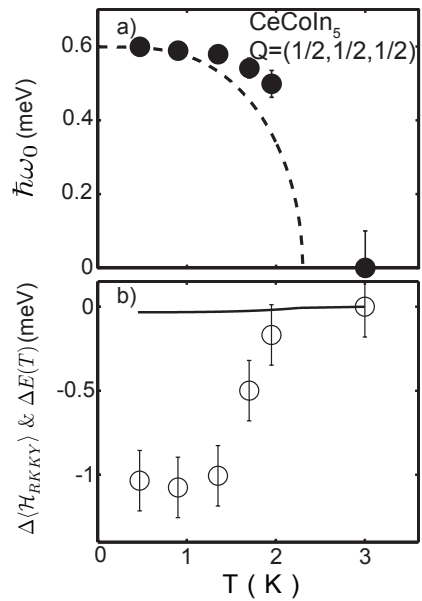}
\caption{(a) Spin resonance energy versus $T$ compared to the scaled d-wave BCS superconducting gap amplitude described text. (b) thermal variation for $T<3$~K of the exchange energy, $\Delta\langle{\cal H}_{RKKY} \rangle$, derived from inelastic neutron scattering compared to the overall electronic energy derived from specific heat data: $\Delta E(T)=\int_0^{T}C(T')dT'-\int_0^{3\ \rm K}C(T')dT'$.}
\label{processed}
\end{figure}

	Two temperature dependent energy scales can be derived from Fig.~\ref{constant_Q}. Fig.~\ref{processed}(a) shows the resonance energy, $\hbar\omega_0$ compared to the temperature dependence of the amplitude of the d-wave SC gap\cite{Yang98:57} scaled to $\hbar\omega_0$ for $T\rightarrow 0$. $\hbar\omega_0$ varies significantly less with temperature than the d-wave superconducting gap amplitude $2\Delta(T)$ calculated by conventional BCS theory indicating that these may be distinct energy scales.

The {\em T}-dependence of the energy of the 4f electron system can be extracted from the corresponding spin susceptibility via the first moment sum-rule\cite{Hohenberg} $
\hbar^2\int_{-\infty}^{\infty}\omega{\rm d}\omega{\cal S}^{\alpha\alpha}({\bf q},\omega)=\frac{1}{2}\langle[[S^{\alpha}_{\bf q},{\cal H}_{4\rm f}],S^{\alpha}_{-{\bf q}}]\rangle
$
While single ion terms produce $\bf q$-independent contributions, Fig. 2 shows strongly $\bf q$-dependent changes in ${\cal S}^{\alpha\alpha}({\bf q},\omega)$ through $T_c$. This indicates that inter-site terms dominate. Neglecting all but isotropic RKKY exchange yields
\begin{equation}
\Delta\langle {\cal H}_{RKKY}\rangle\approx \frac{-\hbar^2}{4\pi(g\mu_B)^2}\int_0^{\infty} \omega
{\rm Tr}\{\Delta\chi^{\prime\prime}({\bf q}_0\omega)\}{\rm d}\omega .
\label{hrkky}
\end{equation}
Here $g=0.86$ is the calculated Land\'{e} factor and we assume ${\bf q}_0\cdot{\bf d}=\pi$ for all displacement vectors, $\bf d$, separating spins with bi-linear exchange interactions. Introducing the cut-off $\hbar\omega_c=1.25$~meV, $\Delta\langle {\cal H}_{RKKY}\rangle$ was calculated from the fit parameters and Eq.~(\ref{hrkky}) and plotted versus $T$ in Fig. 4(b). Also shown is the much smaller overall reduction in electronic energy derived by integrating phonon subtracted specific heat data (solid line).

	The present neutron data associate $\rm CeCoIn_5$ with various magnetic SC with a spin resonance. Most notably, the cuprate superconductors $\rm YBa_2Cu_3O_{6+x}$ and $\rm Bi_2Sr_2CaCu_2O_8$ display a peak in the neutron scattering cross-section near $\hbar\omega$=41 meV which has been associated with formation of a d-wave superconducting ground state.~\cite{Norman00:61,Kao00:61,Morr98:81}  A sharp spin resonance has also been observed in $\rm UPd_2Al_3$ for $\hbar\omega=0.35$~meV$\approx 2.3 k_BT_c$.\cite{Sato01:410} Independent experimental work has established that the SC gap function of all these materials undergoes a sign change $\Delta({\bf q}+{\bf Q_0})=-\Delta({\bf q})$, where $\bf Q_0$ is the wave vector where the spin resonance cross section is greatest. Empirically, a sign change in $\Delta (\bf q)$ thus appears to be a necessary condition for a spin resonance. Assuming line nodes extending along the c-axis, our observation of a spin resonance at ${\bf{Q}_0}=(\frac{1}{2},\frac{1}{2},\frac{1}{2})$ for $\rm CeCoIn_5$ thus indicates $d(x^2-y^2)$ symmetry for the SC order parameter. This result is consistent with recent angular dependent point contact tunnelling data\cite{Rourke05:94,Park05:72,Goll06:383} and theoretical analysis of specific heat data in a magnetic field.\cite{Vorontsov07:unpub}

	Much theoretical work has been devoted towards understanding the spin resonance in high temperature SC. Associating it with transitions to a bound state between collective spin and itinerant electron degrees of freedom, a coherence factor indicates that enhanced resonance intensity at wave vector ${\bf Q}_0$ requires a sign change in $\Delta({\bf q})$ under the corresponding translation in reciprocal space: $\Delta({\bf q}+{\bf Q_0})=-\Delta({\bf q})$. While the bound state should lie below $2\Delta_0$, $\hbar\omega_0$ may depend on the detailed band structure and the coupling constant between AFM and SC. For $\rm Bi_2Sr_2CaCu_2O_{8+\delta}$ ARPES and STM measurements are consistent with $\Delta_0\approx39$~meV\cite{McElroy03:422} while $\hbar\omega_0^{(o)}=42$~meV and $\hbar\omega_0^{(e)}=55$~meV for odd and even parity excitations of the SC bi-layer\cite{Capogna07:75} so that on average $(\hbar\omega_0/2\Delta_0)=0.62$. Recent tunnelling experiments for $\rm CeCoIn_5$ indicate $\Delta_0=0.46$~meV so that $(\hbar\omega_0/2\Delta_0)=0.65$, though we note that considerable debate on the tunnelling spectrum in CeCoIn$_{5}$ remains unresolved.~\cite{Rourke05:94,Park05:72,Goll06:383} The corresponding numbers for $\rm UPd_2Al_3$ are $\Delta_0=0.235$~meV\cite{Jourdan99:398} and $\hbar\omega_0=0.35$~meV\cite{Sato01:410} so that $(\hbar\omega_0/2\Delta_0)=0.74$. The apparent similarity of $(\hbar\omega_0/2\Delta_0)$  over this series of different $d-$wave SC contrasts with $(2\Delta_0/k_BT_c)$, which is 2.8, 4.6, and 6.5 respectively for $\rm UPd_2Al_3$, $\rm CeCoIn_5$, and $\rm Bi_2Sr_2CaCu_2O_{8+\delta}$. 

	The wave vector dependence of the spin resonance may be affected by band structure directly as well as RKKY exchange. Indeed both commensurate and incommensurate spin correlations are observed in compounds closely related to $\rm CeCoIn_5$. The dynamic spin correlation length inferred from the width of constant energy scans is largely unaffected by the transition to SC (see Fig.~\ref{constant_E}) and short-ranged.  It is remarkable that the resonance peak remains well-defined in the absence of long-ranged spatial correlations.  This fact distinguishes $\rm CeCoIn_5$ from $\rm UPd_2Al_3$ which has long range spin order but is similar to the cuprates where a resonance is observed in the absence of long-ranged spin correlations.

	It is interesting to contrast $\rm CeCoIn_5$ with other magnetic SC where a spin resonance has been sought but not found. In $\rm Sr_2RuO_4$ ($0.7~{\rm K}<T_c<1.4~{\rm K}$) funnelling data indicate $2\Delta_0=2.2$~meV and the present interpretation of a range of data indicate a triplet p-wave SC. For a triplet SC a resonance is expected without a sign change in the SC gap function however for $\hbar\omega>0.4$~meV magnetic neutron scattering at ${\bf Q}=(0.7,0.3,0)$ is unaffected by cooling into the SC state.~\cite{Braden02:66} A multi-band model where the nesting bands that contribute to the spin susceptibility are only tangentially involved in SC may provide a resolution.~\cite{Kusunose02:60,Tanatar05:95} Early experiments in $\rm UPt_3$ ($T_c=0.5$~K) also detected no changes in magnetic neutron scattering for $\hbar\omega>0.1$ meV.~\cite{Aeppli88:60} Experiments and theory indicate that the SC gap in $\rm UPt_3$ has line nodes in the basal plane and point nodes along the 6-fold axis however, there is no final resolution regarding the symmetry of the order parameter or the role of magnetic fluctuations in stabilizing it. The apparent absence of a resonance in SC $\rm UPt_3$ should be reexamined.

	We have investigated low-energy magnetic fluctuations in CeCoIn$_{5}$ using neutron inelastic scattering.  A strong spin resonance with an energy of $\hbar\omega_0=0.60\pm0.03$~meV is observed to develop at low temperatures gathering spectral weight from low-energies.  The results indicate strong coupling between spin fluctuations and a d-wave superconductivity.  In addition our analysis of the $T-$dependent spin fluctuation spectrum indicates a large reduction in the RKKY spin exchange energy for $T<T_c$ the relation of which to the development of d-wave superconductivity remains to be understood.

	We acknowledge discussions with P. Coleman, M. Mostovoy, and Z. Tesanovic. Work at JHU was supported by the Natural Science and Engineering Research Council (NSERC) of Canada and the US National Science Foundation through DMR-0306940.  Work carried out at Brookhaven National Labs was supported by the U.S. Department of Energy and Brookhaven Science Associates (DE-Ac02-98CH10886).

\end{document}